\documentclass[12pt, preprint]{emulateapj}
\usepackage[]{natbib}
\shorttitle{Extreme IMFs in UCDs?}
\shortauthors{Mieske \& Kroupa}
\begin{document}

\title{An extreme IMF as an explanation for high M/L ratios in UCD\lowercase{s}? The CO index as a tracer of bottom heavy IMF\lowercase{s}.}

\author{Steffen Mieske\altaffilmark{1}, Pavel Kroupa\altaffilmark{2}}
\altaffiltext{1}{European Southern Observatory,
Karl-Schwarzschild-Strasse 2, 85748 Garching bei M\"unchen, Germany;
{\sf smieske@eso.org}}
\altaffiltext{2}{Argelander-Institut f\"ur Astronomie, Auf dem H\"ugel 71, 53121 Bonn, Germany;  {\sf pavel@astro.uni-bonn.de}} 

\begin{abstract}
  A new type of compact stellar systems, labelled ``ultra-compact
  dwarf galaxies'' (UCDs), was discovered in the last decade.  Recent
  studies show that their dynamical mass-to-light ratios (M/L) tend to
  be too high to be explained by canonical stellar populations, being
  on average about twice as large as those of Galactic globular
  clusters of comparable metallicity. If this offset is caused by dark
  matter in UCDs, it would imply dark matter densities as expected for
  the centers of cuspy dark matter halos, incompatible with cored dark
  matter profiles. Investigating the nature of the high M/L ratios in
  UCDs therefore offers important constraints on the phase space
  properties of dark matter particles.  Here we describe an
  observational method to test whether a bottom-heavy IMF may cause
  the high M/L ratios of UCDs.  We propose to use the CO index at
  2.3$\mu m$ -- which is sensitive to the presence of low-mass stars
  -- to test for a bottom heavy IMF.  In the case that the high M/L
  ratios are caused by a bottom-heavy IMF, we show that the equivalent
  width of the CO index will be up to 30\% weaker in UCDs compared
  to sources with similar metallicity that have canonical IMFs.
  We find that these effects are well detectable with current
  astronomical facilities in a reasonable amount of time (a few hours
  to nights).  Measuring the CO index of UCDs hence appears a
  promising tool to investigate the origin of their high M/L ratios.
\end{abstract}


\section{Introduction}

Within the last decade, a new class of compact stellar systems has
been discovered: the so-called ``ultra-compact dwarf galaxies'' (UCDs,
Phillips {et~al.} 2001; see also Hilker {et~al.} 1999; Drinkwater et
al. 2000 \& 2003; Mieske et al. 2004, 2006, 2007; Ha\c{s}egan et al.
2005).  UCDs are characterised by old stellar populations, absolute
magnitudes $-11>M_V>-13.5$ mag, half-light radii $15<r_h/pc<100$ and
masses $3\times 10^6\lesssim M/M_{\sun} \lesssim 10^8$.  Hence they are
larger, brighter and more massive than the biggest Milky Way globular
clusters (GCs), but at the same time significantly more compact than
typical dwarf galaxies of comparable luminosity. 

There are three main hypotheses on the origin of UCDs: 1. They are
remnant nuclei of tidally stripped dwarf galaxies (Zinnecker et
al.~1988; Bassino { et~al.}  1994; Bekki et~al.  2003). 2. UCDs are
merged star cluster complexes created in galaxy mergers (Fellhauer \&
Kroupa~2002 and 2005). 3. UCDs are genuine compact dwarf galaxies
originating from small-scale peaks in the primordial dark matter power
spectrum (Drinkwater et al. 2004).  In the latter case, UCDs may be
expected to still be dark matter dominated.

\begin{figure}[h]
\plotone{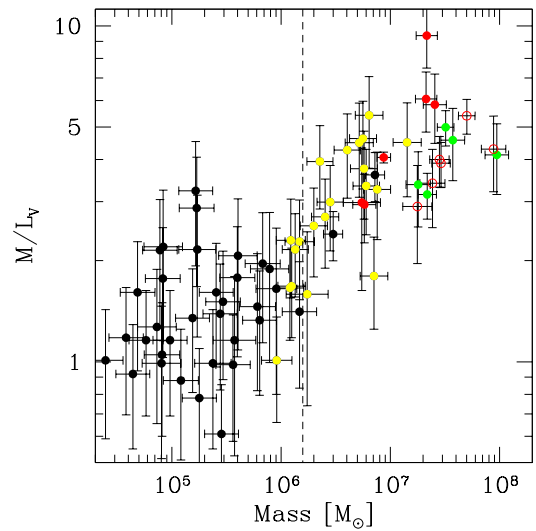}
\caption{This plot shows the mass of compact stellar systems
  plotted vs. their dynamical mass-to-light ratio M/L$_V$. The plot
  covers the mass range from low-mass globular clusters up to the most
  massive UCDs. The vertical dashed line indicates the approximate
  mass 2$\times$10$^6 M_{\sun}$ where the relaxation time equals one Hubble
  time (see text and Dabringhausen et al. 2008). Black dots are Milky
  Way globular clusters from McLaughlin \& van der Marel (2005).
  Filled red circles are Virgo UCDs from Ha\c{s}egan et al. (2005).
  Open red circles are Virgo UCDs from Evstigneeva et al. (2007).
  Yellow filled circles are compact objects in Centaurus A from
  Rejkuba et al. (2007). Filled green dots are Fornax UCDs from Hilker
  et al. (2007). }
\label{mass_ML}
\end{figure}

\section{High M/L ratios of UCD\lowercase{s}}
A powerful way to investigate the nature of UCDs is to measure their
dynamical mass-to-light ratios (M/L). Contrasting the measurements
with predictions from stellar population models, one can estimate the
amount of unseen mass in UCDs, potentially constraining their
  formation scenarios. In the Virgo cluster, Ha\c{s}egan et al.
  (2005) have found several UCDs with M/L ratios well above the ranges
  predicted by stellar population models that assume a canonical IMF
  (Kroupa or Chabrier). Evstigneeva et al. (2007) investigate a
  disjunct sample of 6 Virgo UCDs and find that within the errors,
  their M/L ratios are compatible with canonical SSP models. Hilker et
  al. (2007) study the internal dynamics of 5 Fornax UCDs and the 6
  Virgo UCDs from Evstigneeva et al. (2007). They arrive to the
  conclusion that the UCD M/L ratios tend to be above predictions from
  SSP models assuming a canonical Kroupa or Chabrier IMF, but are
  consistent with a Salpeter IMF, which is characterised by being more
  bottom-heavy than the canonical IMF (Kroupa 2001).

In Fig.~\ref{mass_ML} we plot the available literature M/L$_V$ measurements for
compact stellar systems vs. their mass, ranging from low-mass globular
clusters up to the most massive UCDs.  This sample spans a mass range
of 4 decades.  Clearly, starting at a mass of $\sim$2$\times 10^6$
M$_{\sun}$, the data show a trend of increasing M/L with increasing
mass (this trend has also been noted by Evstigneeva et al. 2007
and Rejkuba et al. 2007).

\begin{figure*}[]
\plottwo{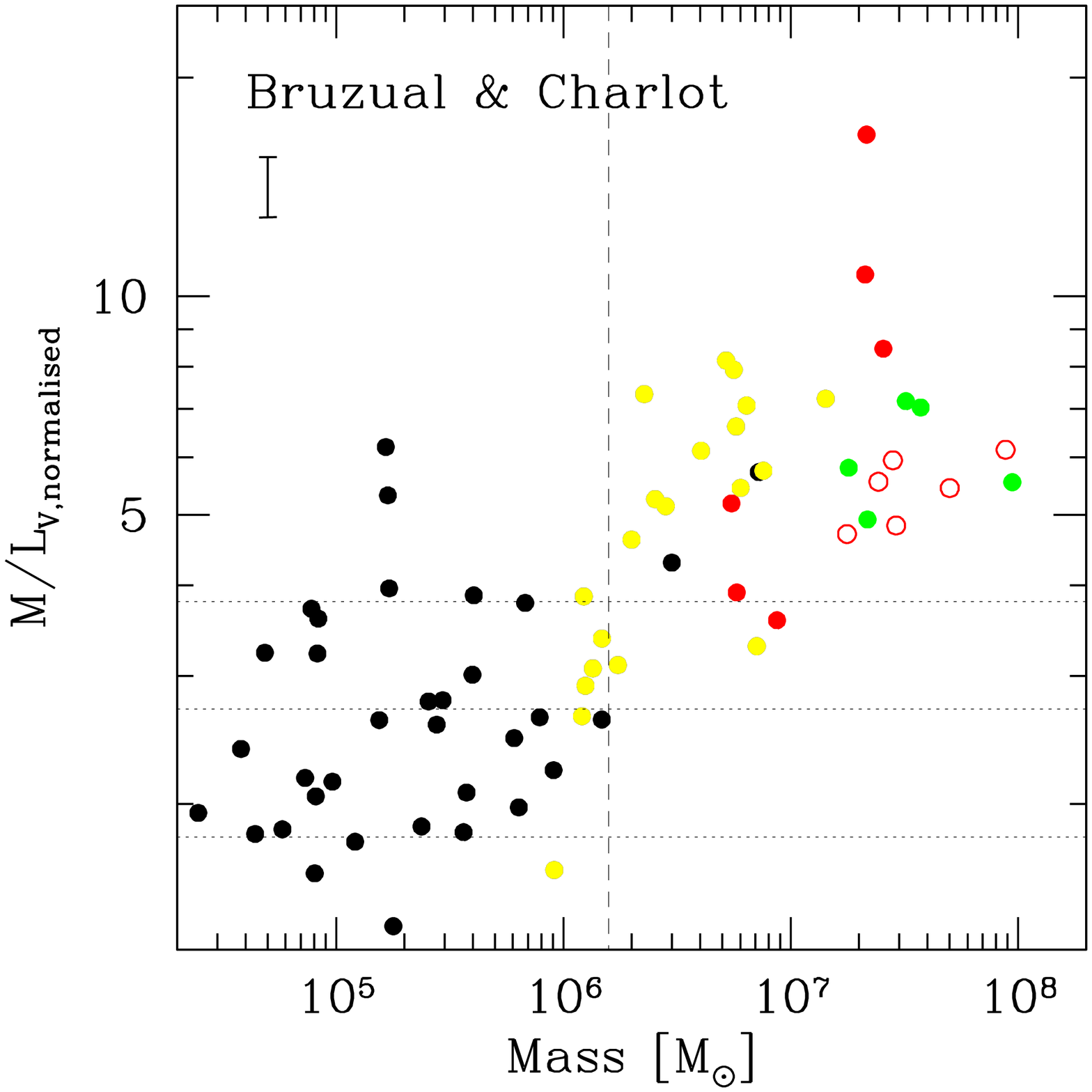}{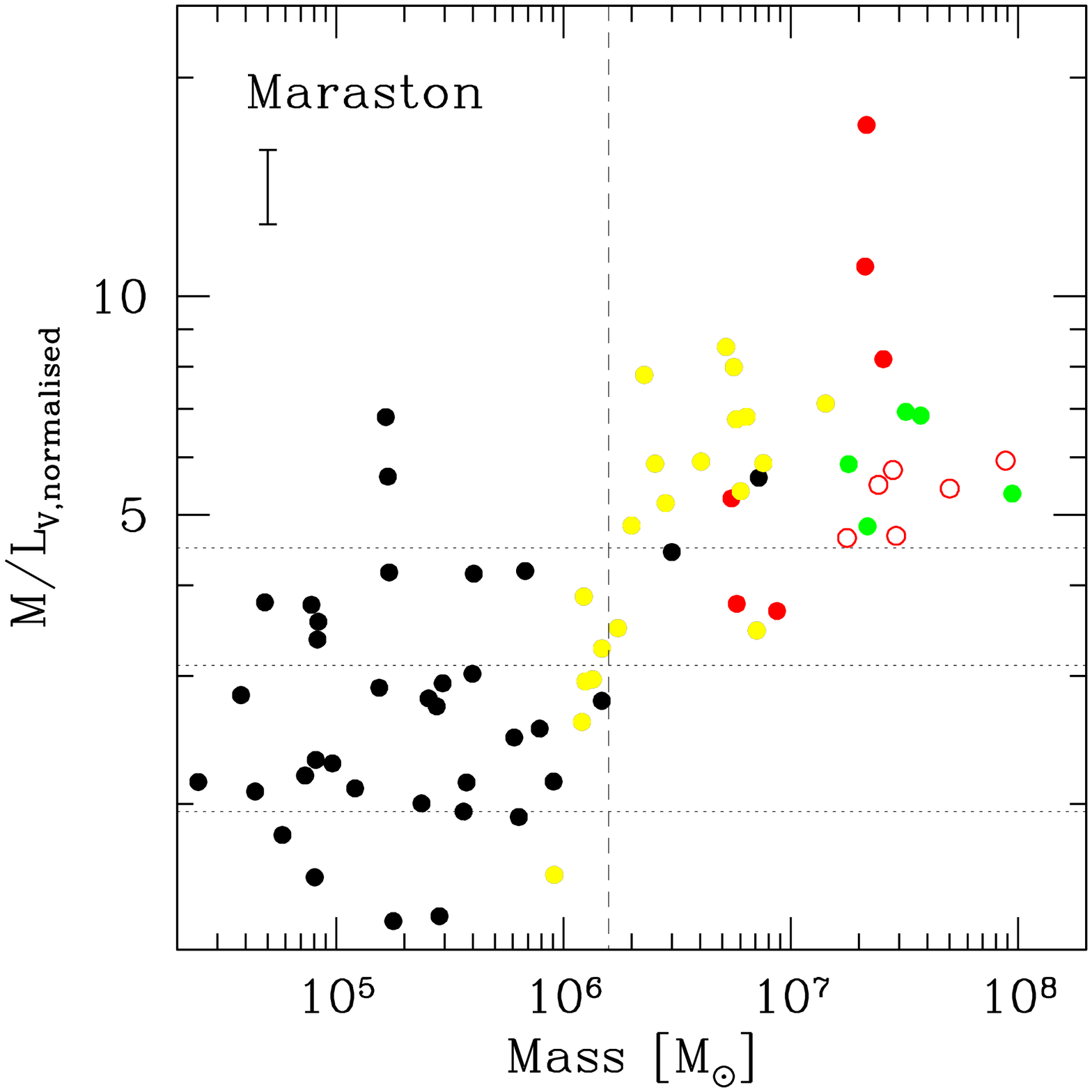}
  \caption{This plot shows the same sample of objects from Fig.~\ref{mass_ML}, but here
    we plot mass vs. the logarithm of M/L ratios {\it normalised to
      solar metallicity}. See text for details on the normalisation.
    The left plot refers to normalisation using the models by Bruzual
    \& Charlot (2003), the right plot refers to normalisation with the
    models of Maraston (2005). The error bar in the top left of each
    plot indicates the average normalisation error assuming a conservative
    uncertainty of 0.3 dex in [Fe/H]. The horizontal lines indicate
    the expected M/L ratio at solar metallicity for ages of 13, 9, and
    5 Gyrs (from top to bottom) in the respective models.  For masses
    $ \ge 2\times10^6 M_{\sun}$ (the realm of UCDs), M/L ratios tend to be
    above the range predicted from the canonical models.}
\label{mass_ML_norm}
\end{figure*}

In this context it is important to stress that stellar population
models predict an increase of M/L$_V$ ratio with increasing
metallicity, as noted previously in several studies (e.g.  Ha\c{s}egan
et al.  2005, Hilker et al. 2007, Evstigneeva et al.  2007).
Therefore, a possible increase of metallicity with mass among the
sample of compact objects investigated will naturally create a trend
of M/L$_V$ with mass as seen in Fig.~\ref{mass_ML}.  In order to
remove such a possible effect from the data, we normalise the M/L$_V$
ratio values in Fig.~\ref{mass_ML} to the same metallicity. The result
of this exercise is shown in Fig.~\ref{mass_ML_norm}. For normalising,
we use the M/L predictions as a function of [Fe/H] for a 13 Gyr
population from the two model sets of Bruzual \& Charlot (2003,
Chabrier IMF) and Maraston (2005, Kroupa IMF) (see also Dabringhausen
et al. 2008). To each of the two sets of model predictions we fit a
function of the form

\begin{equation}
(M/L)_{{\rm theo}}=a+b\times{\rm exp}(c\times[Fe/H])\hspace{0.5cm} \label{form_norm}
\end{equation}

\noindent
 The parameters of the fits are shown in Table 1. For each of the two model sets,
we then normalise the literature M/L values in the following way:

\begin{equation}
(M/L)_{{\rm normalised}}=\frac{(M/L)}{(M/L)_{{\rm theo}}}\times(M/L)_{{\rm theo,0}}\hspace{0.5cm},
\end{equation}

\noindent where $(M/L)_{{\rm theo,0}}$ is the theoretical prediction for [Fe/H]=0, i.e. solar metallicity. 

\begin{table}[]
\caption{Parameters of the exponential fit (see equation \ref{form_norm}) to the relation of M/L as f([Fe/H]) for a 13 Gyr population with a canonical IMF, shown for the two sets of literature models.
\label{table_1}}\vspace{0.2cm}
\small
\begin{center}
\begin{tabular}{|l|rrr|}
\hline\hline Model &  a &  b &
c \\\hline 
Bruzual \& Charlot (2003)&1.98 &1.245 &1.72 \\
Maraston (2005)&3.21 &0.739 &1.29 \\\hline
\end{tabular}\vspace{0.2cm}\\
\end{center}
\vspace{0.5cm} \normalsize
\end{table}

Some details on the metallicities of the literature data: the
values for Galactic GCs are direct measurements, including isochrone
fitting and spectroscopy (McLaughlin \& van der Marel 2005). The [Fe/H]
values for the CenA sources from Rejkuba et al.  (2007) are derived from
their (V-I) and (B-V) colours. For this we adopted the mean [Fe/H]
derived from applying the calibration relations from Kissler-Patig et
al.  (1998) for (V-I) and from Barmby et al. (2000) for (B-V).  The
[Fe/H] values from Evstigneeva et al.  (2007) are derived from their
measurement of the MgFe Lick line index, applying the corresponding
calibration relation used in Mieske et al.  (2007). The [Fe/H] values
for the sources from Ha\c{s}egan et al.  (2005) are taken from their
paper, and are based on broad-band colours. For 3 out of 5 Fornax UCDs
from Hilker et al.  (2007), spectroscopic [Fe/H] estimates from Mieske
et al. (2006) are used. For the 2 remaining UCDs, [Fe/H] is estimated
based upon their (V-I) colours using Kissler-Patig et al. (1998). In Table 2
we list the adopted [Fe/H], M/L and M/L$_{\rm norm}$ for the
compact objects in CenA, Fornax and Virgo.

\begin{table*}[]
\renewcommand{\baselinestretch}{1.00}
\caption{Literature M/L ratios, metallicities [Fe/H], and normalised M/L$_{\rm norm}$ for compact objects in CenA, Virgo and Fornax (Fig.~\ref{mass_ML_norm}). The M/L ratios are normalised to solar metallicity (see text), using Bruzual \& Charlot (2003, col. 4) and Maraston (2005, col. 5).\label{table_2}}
\small
\begin{center}

\begin{tabular}{|l|r|r|r|r|r|}
\hline\hline Source & M/L &  [Fe/H] & (M/L)$_{\rm norm,BC}$ & (M/L)$_{\rm norm,M}$ 
& Environment\\
& & & (M/L)$_{\rm sol}=3.7$ & (M/L)$_{\rm sol}=4.5$  &\\\hline 
HGHH92-C7 & 3.26 &  $-$1.30 &  5.75 &  5.89 & CenA$^1$\\
HGHH92-C11 & 5.42 & $-$0.46 &  7.07 &  6.82 & CenA$^1$\\
HHH86-C15=R01-226 & 2.28 & $-$0.75 &  3.38 &  3.27 & CenA$^1$\\
HGHH92-C17 & 3.76 &  $-$1.30 &  6.62 &  6.77 & CenA$^1$\\
HGHH92-C21 & 4.62 &  $-$1.20 &  7.92 &  7.99 & CenA$^1$\\
HGHH92-C22 & 2.99 &  $-$1.20 &  5.14 &  5.19 & CenA$^1$\\
HGHH92-C23 & 1.80 &  $-$1.50 &   3.30 &  3.46 & CenA$^1$\\
HGHH92-C29 & 4.27 & $-$0.67 &  6.12 &  5.92 & CenA$^1$\\
HGHH92-C36=R01-113 & 2.54 &  $-$1.50 &  4.62 &  4.83 & CenA$^1$\\
HGHH92-C37=R01-116 & 1.66 & $-$0.95 &  2.64 &  2.59 & CenA$^1$\\
HHH86-C38=R01-123 & 1.68 &  $-$1.20 &  2.91 &  2.95 & CenA$^1$\\
HGHH92-C41 & 2.16 & $-$0.65 &  3.07 &  2.97 & CenA$^1$\\
HGHH92-C44 & 3.95 &  $-$1.60 &  7.33 &  7.79 & CenA$^1$\\
HCH99-2 & 4.49 &  $-$1.50 &  8.15 &  8.51 & CenA$^1$\\
HCH99-15 & 3.34 &    $-$1.00 &  5.45 &  5.39 & CenA$^1$\\
HCH99-16 & 2.70 &  $-$1.90 &  5.25 &  5.88 & CenA$^1$\\
HCH99-18 & 4.50 & $-$0.98 &  7.23 &  7.12 & CenA$^1$\\
HCH99-21 & 1.59 &    $-$2.00 &   3.10 &  3.49 & CenA$^1$\\
R01-223 & 2.30 &  $-$1.10 &  3.86 &  3.86 & CenA$^1$\\
R01-261 & 1.01 & $-$0.98 &  1.62 &   1.60 & CenA$^1$\\
S314 & 2.94 &  $-$0.50 &  3.91 &  3.77 & Virgo$^2$\\
S417 & 5.83 &  $-$0.70 &  8.46 &  8.19 & Virgo$^2$\\
S490 & 4.06 &  0.18 &  3.58 &  3.69 & Virgo$^2$\\
S928 & 6.06 &  $-$1.30 &  10.7 &    11.0 & Virgo$^2$\\
S999 & 9.36 &  $-$1.40 &  16.7 &  17.2 & Virgo$^2$\\
H8005 & 2.98 &  $-$1.30 &  5.18 &  5.27 & Virgo$^2$\\
VUCD1 & 4.00 & $-$0.76 &  5.94 &  5.76 & Virgo$^3$\\
VUCD3 & 5.40 & $-$0.01 &  5.44 &  5.43 & Virgo$^3$\\
VUCD4 & 3.40 &    $-$1.00 &  5.55 &   5.50 & Virgo$^3$\\
VUCD5 & 3.90 & $-$0.36 &  4.83 &  4.68 & Virgo$^3$\\
VUCD6 & 2.90 &    $-$1.00 &   4.70 &  4.65 & Virgo$^3$\\
VUCD7 & 4.30 & $-$0.66 &  6.14 &  5.94 & Virgo$^3$\\
UCD1 & 4.99 & $-$0.67 &  7.17 &  6.93 & Fornax$^4$\\
UCD2 & 3.15 &  $-$0.90 &  4.93 &  4.82 & Fornax$^4$\\
UCD3 & 4.13 & $-$0.52 &  5.54 &  5.35 & Fornax$^4$\\
UCD4 & 4.57 & $-$0.85 &  7.02 &  6.85 & Fornax$^4$\\
UCD5 & 3.37 &  $-$1.20 &   5.80 &  5.87 & Fornax$^4$\\\hline
\end{tabular}\vspace{0.2cm}\\
\begin{footnotesize}
$^1$ Rejkuba et al. (2007). [Fe/H] from broad-band colours, see text.

$^2$ Ha\c{s}egan et al. (2005). [Fe/H] from broad-band colours, see text.

$^3$ Evstigneevae et al. (2007). [Fe/H] from spectroscopy, see text.

$^4$ Hilker et al. (2007). [Fe/H] from spectroscopy
for UCD1, UCD3, UCD4, from (V-I) for UCD2 and UCD5.

\end{footnotesize}
\end{center}
\vspace{0.5cm} \normalsize
\renewcommand{\baselinestretch}{1.0}
\end{table*}

In Fig.~\ref{mass_ML_norm} we show the normalised M/L ratios, in
separate plots for the two model sets chosen for normalisation. We note
that the normalisation to solar metallicity is of course an arbitrary
choice, but we prefer this particular choice in order to have a well
defined reference that is independent on the actual metallicity of the
investigated population. In Fig.~\ref{mass_ML_norm}, a change of the
normalisation would change nothing but the scale of the y-axis. The
location of the data with respect to the M/L ratios expected from the
models is independent on the choice of normalisation. 

For both model choices plotted in Fig.~\ref{mass_ML_norm} it is clear
that the distribution of mass vs. M/L ratios does exhibit a very
interesting separation at about 2$\times$10$^6$ M$_{\sun}$: sources
with lower masses have M/L ratios in line with canonical stellar
populations, whereas sources with higher masses have M/L ratios that
tend to be above those for canonical stellar populations. The ratio of
the average M/L ratio between low- and high-mass sources is 2.3 $\pm$
0.2 for both choices of normalisation. The occurence of this
break at 2$\times$10$^6$ M$_{\sun}$ is in line with previous findings
of breaks in terms of size (Ha\c{s}egan et al.  2005, Mieske et al.
2006) and metallicity (Mieske et al. 2006) at about the same mass.  It
hence strengthens the proposed separation between canonical globular
clusters and ultra-compact dwarf galaxies at this mass (Ha\c{s}egan et
al.  2005, Mieske et al. 2006).  Interestingly, the separation
  mass corresponds to the mass range where the relaxation time of compact
  stellar systems is comparable to a Hubble time
  (Dabringhausen et al.  2008).

The high M/L ratios for UCDs can be explained by the presence of dark
matter, which would have to be at very high densities ($\ge 10
M_{\sun}$ pc$^{-3}$ within 20 pc). This is comparable to central
density values expected for cuspy dark matter halos (Walker et al.
2007). However, it is incompatible with recent claims of cored dark
matter halos with low central densities in dwarf spheroidal galaxies
(e.g.  Gilmore et al.  2007), which are the most dark matter-dominated
systems known (see however Metz \& Kroupa~2007 for an entirely
different interpretation).  Investigating the nature of the high M/L
ratios in UCDs therefore offers important constraints on the phase
space properties of dark matter particles.  \vspace{0.15cm}

\section{An extreme IMF as the reason for high M/L?}

Here, we describe a method to test whether a bottom heavy stellar initial mass
function (IMF) causes the high M/L$_V$ ratios of UCDs. Such steep
low-mass stellar IMFs may result if the radiation field, stellar winds
and supernova explosions in UCDs were so intense during their
formation, that pre-stellar cloud cores were ablated before they could
fully condense to stars (Kroupa \& Bouvier 2003).  By confirming such
an overabundance of low-mass stars with respect to a canonical IMF
(Kroupa 2001), we would for the very first time have unambiguous
evidence for a radically different star-formation process under
extreme physical conditions when UCDs formed.

By excluding this scenario we would know that the high M/L$_V$ ratios
are probably due to dark matter. This could be in the form of stellar
remnants, perhaps as a result of a top-heavy IMF in primordial and
pristine star formation regions (POP III, Jimenez \& Haiman 2006,
Tornatore et al.  2007, Dabringhausen et al. 2008).  Alternatively, it
could be due to non-baryonic dark matter, which in turn would open new
avenues towards studying the phase-space properties of dark matter
particles (see above).  We note that tidal heating (Fellhauer \&
Kroupa 2006) may increase the measured M/L ratio of individual UCDs
after a very close passage ($d<500$pc) next to the host galaxy's
center. However, this can certainly not explain the existence of the
general and very clear trend of M/L with mass
(Fig.~\ref{mass_ML_norm}).\vspace{0.15cm}

\begin{figure*}[]
\begin{center}
  \plottwo{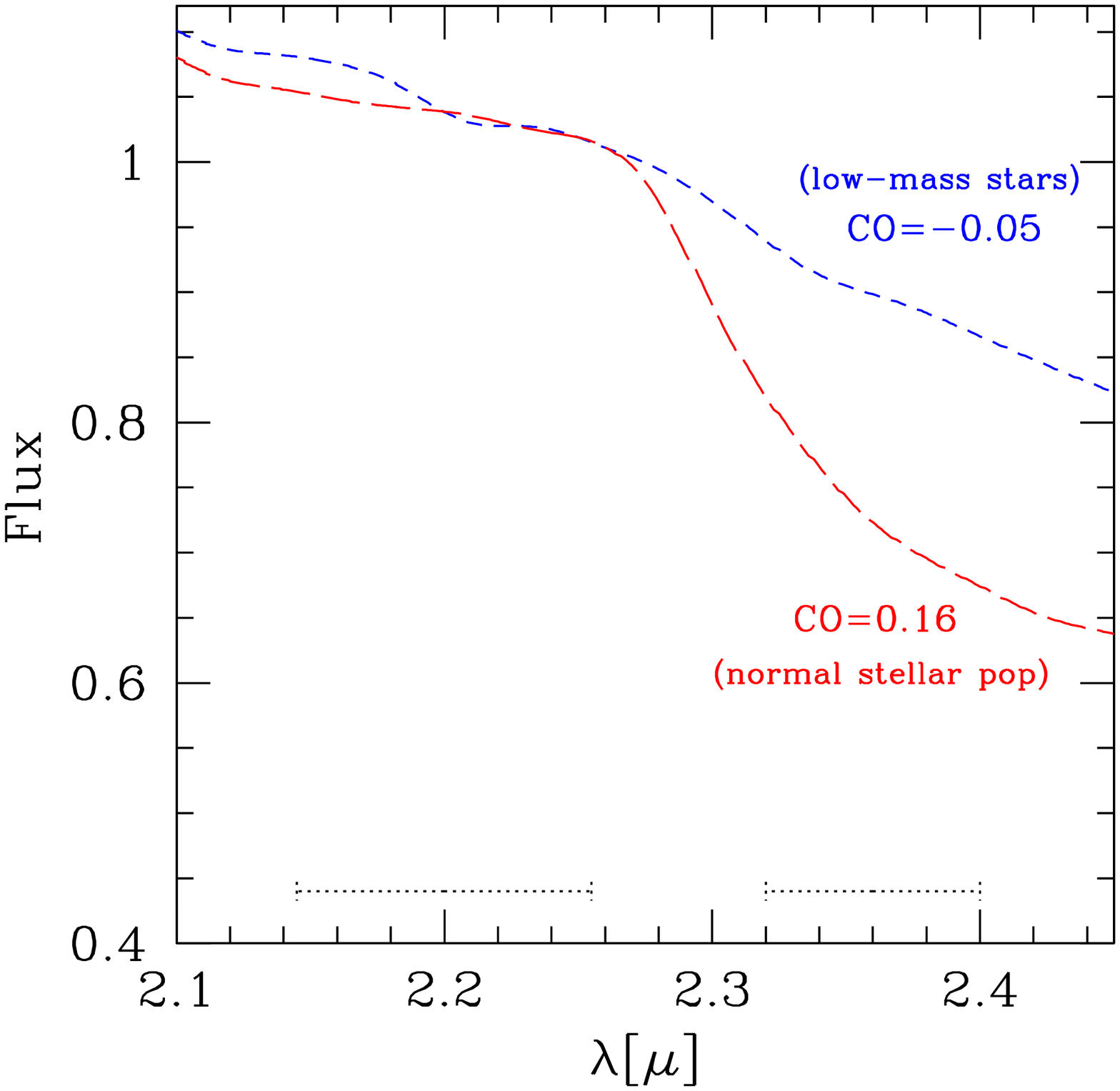}{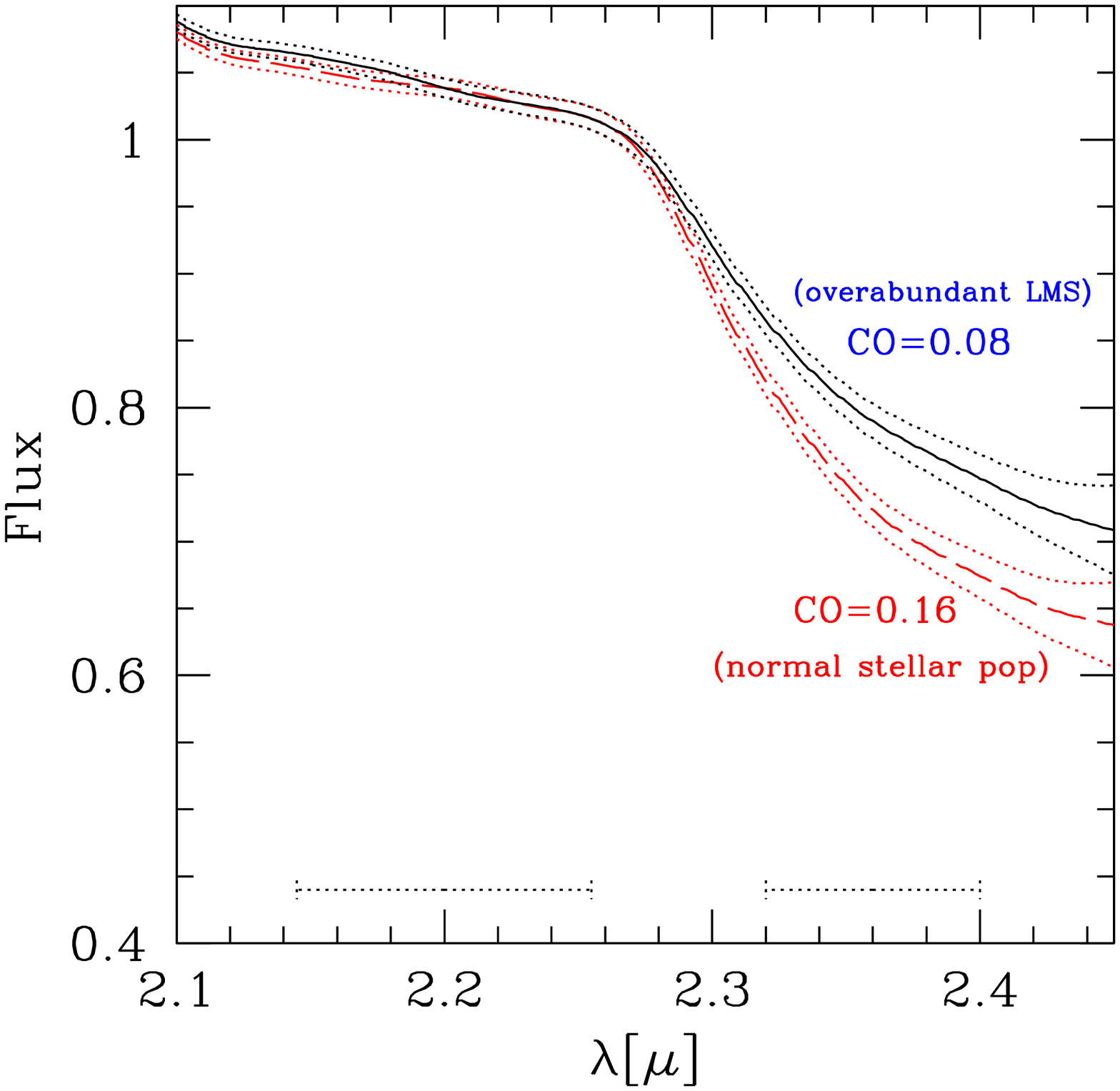}
  \caption{These plots illustrate how an overabundance of
    low-mass stars in an old stellar population can be estimated from
    spectroscopy around the CO band (Kroupa \& Gilmore 1994). The
    feature and continuum band definition of the photometric index is
    indicated by dotted lines at the bottom (Frogel et al. 1978).
    {\bf Left:} Shown is a NIR spectrum of a M2II giant star (lower
    long dashed curve) and a M2V dwarf star (upper short dashed curve)
    from the catalog of Lancon \& Rocca-Volmerange (1992), smoothed to
    0.04$\mu$ resolution.  The CO index 0.16 mag of the M giant
    corresponds to the typical CO index of old stellar populations of
    intermediate metallicity ([Fe/H]$\sim-$0.7 dex, e.g.  Frogel et
    al. 1978 and 2001, Goldader et al.  1997, Ivanov et al.  2000).
    The M dwarf has a much weaker CO feature, representative of a
    population of pure low-mass stars with upper mass cutoff $m_{\rm
      cut}\simeq 0.5$ M$_{\rm sun}$ (Kroupa \& Gilmore 1994).  Under
    the hypothesis that the high M/L ratios of UCDs are caused by an
    overabundance of unevolved low-mass stars like the M dwarf (Kroupa
    \& Gilmore 1994), one would require those stars to contribute a
    significant fraction to the total UCD mass (3/4 for the highest
    M/L UCD known), and also a certain fraction to the K-band
    luminosity (see text).  {\bf Right:} The spectrum of the canonical
    stellar population is shown as in the left panel. The solid line
    now indicates the case of 40\% K-band luminosity contribution from
    an additional population of low-mass stars. This luminosity
    fraction is representative for the case when the additional
    low-mass stars make up 3/4 of the total mass, at a metallicity of
    about [Fe/H]=$-$0.7 dex (see text and Fig.~\ref{coindex}). The
    dotted lines indicate the 1 $\sigma$ error range of the spectrum,
    assuming a S/N of 150 per 0.04$\mu$ resolution element. For
    calculating the S/N as a function of wavelength, we assumed the
    sky brightness as given in the Exposure Time Calculator for the
    VLT ISAAC instrument (http://www.eso.org/observing/etc/).}
\label{cospectra}
\end{center}
\end{figure*}

\subsection{The CO index as a tracer of a bottom-heavy stellar IMF}
\vspace{0.05cm}

\begin{figure*}[]
\begin{center}
  \plottwo{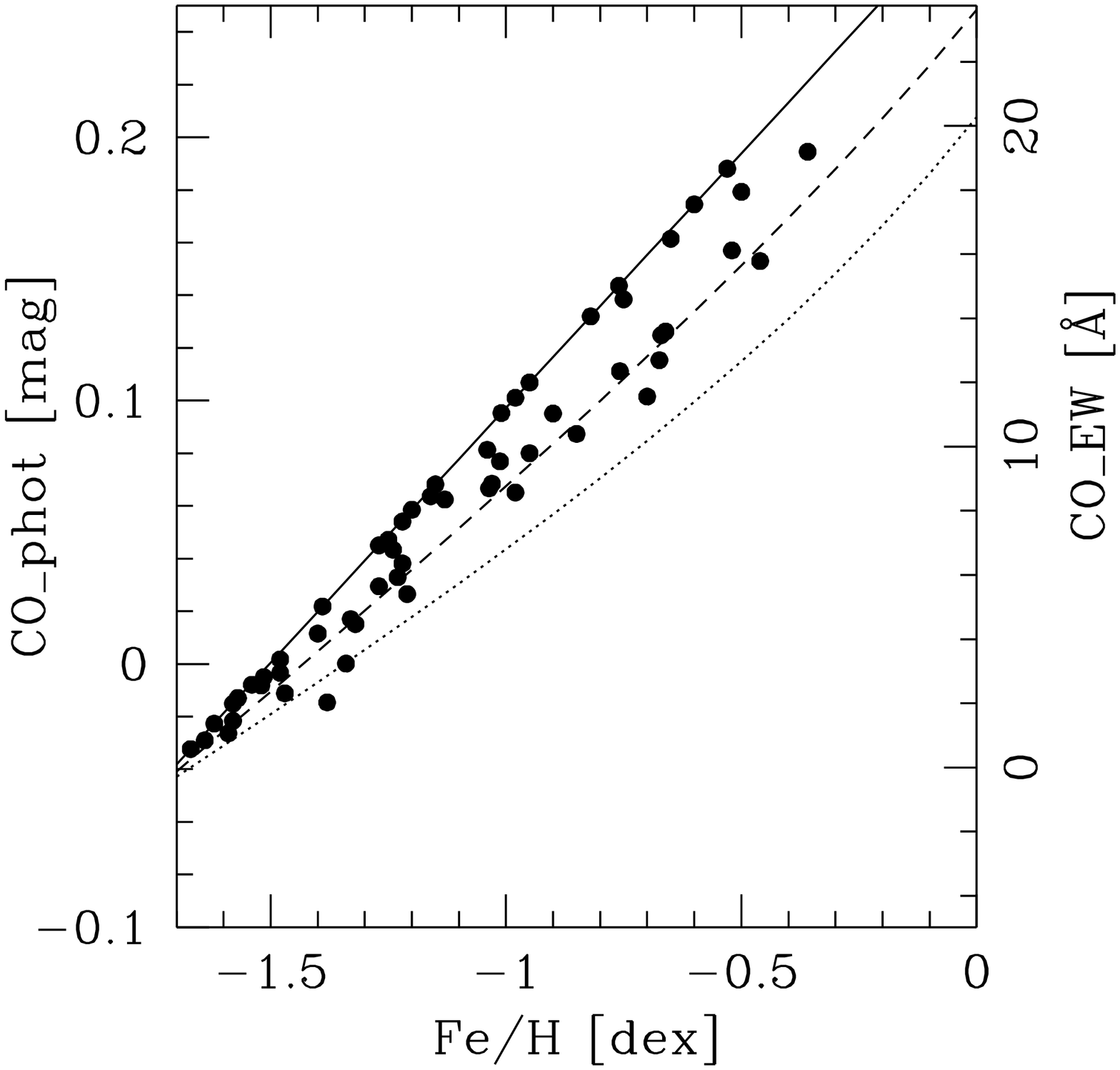}{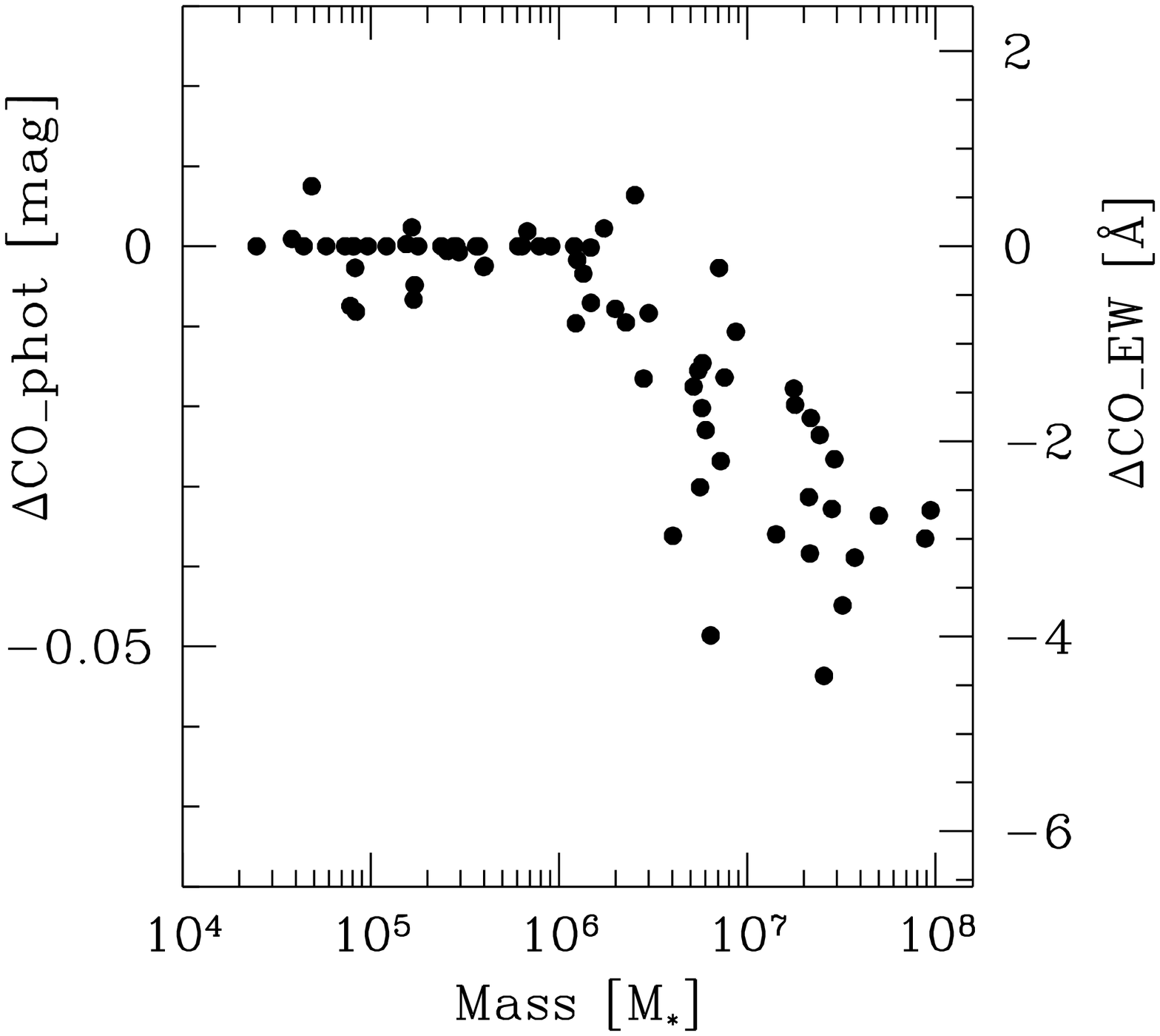}
  \caption{{\bf Left}: Expected CO index for the compact stellar
    systems shown in Fig.~\ref{mass_ML} plotted vs. their
    [Fe/H] values.  The CO indices are calculated based on the
    assumption that an increase in the M/L ratio above the average value found
    for Galactic GCs (see text) is caused by an overabundance of
    low-mass stars. We plot both the photometric CO index and
    spectroscopic CO equivalent width, which are linked by a well
    defined linear relation (see text). The solid line indicates the
    calibration between CO and [Fe/H] from Frogel et al.  (2001),
    using Galactic GCs. As an aid for the eye we also indicate the offset
    with respect to the Frogel et al.  relation, assuming two
    different fractions of additional low-mass stars. The dashed curve
    corresponds to a mass fraction of 55\% contributed by additional
    low-mass stars, representative of the average M/L ratio offset
    between UCDs and GCs in Fig.~\ref{mass_ML}. The dotted curve
    indicates a mass fraction of 75\%, corresponding to the most
    extreme observed UCDs. {\bf Right}: the expected offset from the
    Frogel et al. [Fe/H]$-$CO relation from the left panel is plotted
    vs. the mass of the compact stellar systems. }
\label{coindex}
\end{center}
\end{figure*}

For diagnosing a bottom heavy IMF, we propose to study a portion of
the spectrum where the hypothetical overabundant population of
low-mass (main sequence) stars contributes significantly to the
integrated spectrum. In this region, the shape of the spectral energy
distribution (SED) emitted by these unevolved dwarf stars must be
different to that of the evolved giant stars.  This is the case for
the near-infrared wavelength region of the CO band ($2.3 \mu < \lambda
< 2.42 \mu$, see Fig.~\ref{cospectra}).  For intermediate to high
metallicities, late type giant stars have a very strong CO absorption
feature, while late type dwarf stars have a much weaker feature (see
Figs.~\ref{cospectra} and~\ref{coindex}), independent of metallicity
(e.g. Frogel et al.  1978).  At a given metallicity, the strength of
the CO absorption feature anti-correlates with the fraction of
low-mass stars present in a stellar population. One thus expects a
weaker CO index for high M/L ratio sources, if the high M/L ratios are
caused by low-mass stars.

Our working hypothesis is that the stellar population of a high M/L
ratio UCD consists of a canonical part with a 'normal' M/L and IMF,
{\it plus} an additional population of low mass stars. Kroupa \&
Gilmore (1994) present a detailed analysis of such a scenario. They
quantify which luminosity fraction $f_{\rm lms,K}$ of the composite
overall luminosity is contributed by this additional population of
low-mass stars in the CO wavelength region, which is close to the
K-band. In the context of this paper this fraction can be written in terms of luminosities as

\begin{equation}
f_{\rm lms,K}= \frac{L_{\rm lms,K}}{L_{\rm lms,K}+L_{\rm
    UCD,normal,K}} \hspace{0.5cm}
\end{equation}

 Obviously, $f_{\rm lms,K}=0$ for a normal
stellar population without additional low-mass stars. In terms of
masses and M/L ratios we re-write 
\begin{tiny}
{\begin{equation}
f_{\rm lms,K}=
\frac{(M/L_K)^{-1}_{\rm lms}\times M_{\rm lms}}{(M/L_K)^{-1}_{\rm lms}\times M_{\rm lms}
  + (M/L_K)^{-1}_{\rm UCD,normal}\times M_{\rm UCD,normal}} 
\end{equation}
}\end{tiny}

The term $(M/L_K)_{\rm UCD,normal}$ ranges between 1.4 and 0.66 for [Fe/H] between -2 and 0 dex, assuming a 12 Gyr population (Worthey et
al. 1994).  Introducing the ratio $r= \frac{M_{\rm lms}}{M_{\rm UCD,normal}}$, we get

\begin{equation}
f_{\rm lms,K}= \frac{(M/L_K)^{-1}_{\rm
    lms}\times r}{(M/L_K)^{-1}_{\rm lms}\times r + (M/L_K)^{-1}_{\rm UCD,normal}}\hspace{0.5cm} 
\end{equation}

The ratio $r$ can be expressed as $r=\frac{(M/L_V)_{\rm
    UCD}-(M/L_V)_{\rm normal}}{(M/L_V)_{\rm normal}}$. As
$(M/L_V)_{\rm normal}$ we assume the mean normalised M/L ratio of
compact stellar system less massive than 2$\times$10$^6$ M$_{\sun}$ (see
Fig.~\ref{mass_ML}), which is 2.75. For example, we get $r=5.1$ for the
highest $M/L_V$ ratio UCD (object S999 with normalised $M/L_V$=17).
The term $(M/L_K)_{\rm lms}$ in $f_{\rm lms,K}$ depends on the slope
$\alpha$ and cutoff mass $m_{\rm cut}$ in the mass distribution of the
additional low-mass-star population. From Kroupa \& Gilmore (1994) we
find a range of $ 4.5 < (M/L_K)_{\rm lms} < 9$ for the parameter space
$0.5<m_{\rm cut}/M_{\sun}<0.7$ and $0<\alpha<1.5$.

We can now express the CO index as a function of $f_{\rm lms,K}$, and
hence as a function of $M/L_V$,

\begin{equation}
CO=(1-f_{\rm lms,K}) \times CO_{\rm canonical} + f_{\rm lms,K} \times CO_{\rm lms}\hspace{0.5cm}
\end{equation}

From Fig.~\ref{cospectra} and Kroupa \& Gilmore (1994) we adopt
$CO_{\rm lms}=-0.05$ mag. The choice of $CO_{\rm canonical}$ requires
a brief discussion. The CO index of old stellar populations with
canonical IMF is dominated by giant stars. Since the dominant stellar
type in the integrated light changes towards earlier types for lower
metallicitiy, and because the CO index anticorrelates with surface
temperature, the CO index anticorrelates with integrated metallicity
(e.g. Frogel et al. 2001). We use the linear calibration between
[Fe/H] and CO by Frogel et al. (2001) (see also Fig.~\ref{coindex}) derived from Galactic globular clusters in order to
determine $CO_{\rm canonical}$ at a given [Fe/H].

\vspace{0.15cm}

For each of the compact stellar systems from Fig.~\ref{mass_ML_norm}, we
can now calculate the expected CO index, under the hypothesis that a M/L
ratio offset above the average value for Galactic globular clusters is
explained by an additional population of low-mass stars. For this we
assume the mean of the CO indices corresponding to the two extreme
values of $(M/L_K)_{\rm lms}$ from above. The result of this exercise
is shown in Fig.~\ref{coindex}.  In the left panel we plot [Fe/H]
against the expected CO index. We show both the photometric index
(Fig.~\ref{cospectra}) and the spectroscopic one (feature band between
2.291 and 2.302$\mu${, see Frogel et al. 2001 for the exact pass-band
definitions}), using the well defined linear relation between both
indices (Frogel et al. 2001). We overplot the calibration relation
between [Fe/H] and CO from Frogel et al. (2001). 

There can be
substantial offsets $\gtrsim 0.05$ mag due to the higher M/L ratios,
corresponding to 30\% of the equivalent with of the CO absorption
feature. As aid for the eye we also indicate in Fig.~\ref{cospectra}
the offset relation for two different fractions of additional low-mass
stars.  The dashed curve corresponds to a mass fraction of 55\%
contributed by additional low-mass stars, which is representative of
the average M/L ratio offset between UCDs and GCs in
Fig.~\ref{mass_ML}. The dotted curve indicates a mass fraction of
75\%, corresponding to the most extreme observed UCDs.  In the right
panel of Fig.~\ref{coindex}, we plot the expected offset from the
Frogel et al. [Fe/H]$-$CO relation vs. mass of the compact stellar
systems. The largest offsets are expected for high masses, given the
increase of M/L ratios in the UCD regime.

In the next
section we assess the observational detectability of these offsets.

\begin{figure}[]
\begin{center}
  \plotone{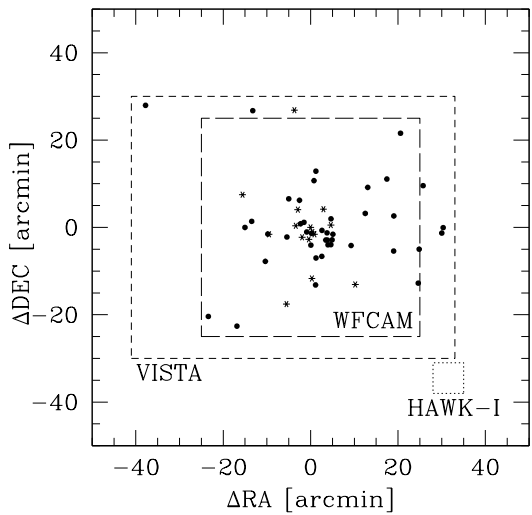}
  \caption{Positions of UCDs in Fornax (solid dots) and Virgo (asterisks) relative to the central galaxies in either cluster, NGC 1399 and M87. The FOV of HAWK-I, VISTA/VIRCAM and WFCAM/UKIRT is indicated.}
\label{map}
\end{center}
\end{figure}

\subsection{Observability of the change in CO as f(M/L)}

Looking at Fig.~\ref{coindex}, there are three minimum requirements
for detecting a drop in the CO index {\it in a single source} due to a
high M/L ratio. First, one requires an accurate calibration between CO
and [Fe/H], applicable to $\sim$0.1 dex precision. Second, one
requires an [Fe/H] measurement of a single source accurate to
$\sim$0.1 dex.  Third, one requires measurement of the CO index to an
accuracy of 0.01$-$0.02 mag, or 5-10\% in terms of equivalent width.
We note that the existing calibration between CO and [Fe/H] by Frogel
et al.  (2001) is performed with Galactic GCs, and has a
cluster-to-cluster dispersion of the order $\sim$0.15 dex. It is
reasonable to assume that future calibration attempts including
population synthesis models (e.g. Brodie et al. 2007) will reach
accuracy levels down to $\sim$0.10 dex. Still, even with an accurate
calibration, an offset in a single object measurement may be detected
with very accurately measured CO and [Fe/H] only for the largest expected
CO offsets.

The more efficient way to detect a change in CO are relative
measurements of statistically significant samples ($\ge$10$-$20
objects) covering a broad range in metallicities.  Sources with M/L
ratios consistent with canonical IMFs have to be observed in parallel
to UCDs with high M/L ratios. The aim is to arrive at a comprehensive
sample of precise measurements of M/L, CO and [Fe/H], obtained with
identical or similar instrument settings.  This changes the
requirements on [Fe/H] from $\sim$0.1 dex accuracy to $\sim$0.1 to 0.2
dex relative precision. This is well possible with moderately high S/N
optical spectroscopic measurements of [Fe/H] (e.g. Mieske et al.
2006), requiring on-source integration times of a few hours at 6-10m
class telescopes.  To cover a substantial portion of
the known UCD population, only a few pointings of multi-object
spectroscopy with moderately large FOV (0.25 to 0.5 degree) would be
required. See a map of all known Fornax and Virgo UCDs in
Fig.~\ref{map}. That is, within a
few nights of observations, the required
precision of [Fe/H] can be reached for the known UCD population.

What about the necessary integration time for precisely measuring the
CO feature depth? In the right panel of Fig.~\ref{cospectra} we show a
smoothed NIR spectrum around the CO regions, indicating the 1$\sigma$
envelope for a S/N=150 per 0.04$\mu$ resolution element at 2.2$\mu$.
Integrating over the feature and continuum regions, the precision of
measuring the CO index is about 0.01 mag.  To calculate the
integration time needed to achieve this S/N, we use the Exposure Time
Calculator (ETC) of ISAAC on the VLT (Version 3.2.1). We assume
V=20.0 mag as object magnitude, which is at the faint end of the
observed UCD magnitude range in Fornax and Virgo, and thus gives an
upper limit on the required integration time.  As input for the ISAAC
ETC we adopt $(V-K)$=2.5 (Worthey 1994 for a 12 Gyr population of
[Fe/H]=-0.7 dex), hence K=17.5 mag. Assuming a K-band seeing of
0.7$''$, a slit-width of 1$''$, a black-body curve of 4000K, and the
low-resolution spectroscopy mode, we require $\sim$7 hours on-source
integration time to achieve S/N=150 per 0.04$\mu$ at 2.2$\mu$.  As
comparison objects with canonical M/L ratios, one will want to choose
{\it bright} objects, requiring much shorter integration times. Those
could be massive GCs in more nearby galaxies, for which M/L ratios and
metallicities are available (e.g.  GCs in NGC 5128, see Rejkuba et al.
2007).

The ideal way to measure the CO index for a comprehensive sample of
UCDs certainly is wide-field imaging with purpose-build narrow-band
(NB) filters covering the CO feature and continuum, each of width
$\sim$0.1 $\mu$ (Fig.~\ref{cospectra}). In Fig.~\ref{map} we
indicate the projected distribution of UCDs in the Fornax and Virgo
cluster relative to the respective central cluster galaxies. For
comparison we show the FOV of available IR imagers on 4-8m class
telescopes. This plot shows that most of the UCDs in either cluster
are contained within one WFCAM or VISTA pointing.
For the above example of a UCD with K=17.5 mag, we can estimate the
required amount of integration time necessary to achieve S/N=100 for a
point-source image in the CO continuum and feature pass-band.  For
this we use the VISTA ETC (http://www.ast.cam.ac.uk/vdfs/etc) and
calculate the required time for S/N=100 with the Ks filter, which is
centred on 2.15$\mu$ with FWHM=0.3$\mu$.  For K=17.5 mag, the
required integration time using this Ks filter is about 5400 seconds,
or 1.5 hours. To convert this into the time required for the same S/N
with a NB CO filter, two things need to be taken into account. First,
the CO feature bands are 3 times narrower than the Ks filter. Second,
the sky brightness in the CO feature band (2.35$\mu$) is about 2.5
times higher than at 2.15$\mu$ (see ISAAC ETC). Taking these two
differences into account, one requires on-source integration of 5.6
hours for the CO continuum band and 27 hours for the CO feature band.

\section{Conclusions}
We conclude that measuring the depth of the CO feature (2.35$\mu$) in
the near-IR SED of old extragalactic systems is a promising tool
to constrain the low-mass slope of the IMF, provided that precise
metallicity estimates ($\sim$0.15 dex) are available. For the case of
UCDs, such observations offer an important test as to whether their
high M/L ratios are caused by very bottom heavy IMFs. Given that an
alternative explanation for these high M/L ratios is dark matter at
very high densities, the study of the IMF in UCDs will allow important
constraints on the phase space properties of dark matter particles.

To demonstrate the feasibility of the method proposed here, the first
step is to obtain low resolution spectroscopy of the UCDs with largest
expected drop in CO, and reference sources of comparable metallicity
with normal M/L ratios.  As a next step, wide-field NIR imaging with
appropriate narrow-band filters will allow to probe large
representative samples of UCDs within a few nights of telescope time.
With such observations it will also be possible to investigate 
spatial variations in the IMF in early-type giant and dwarf galaxies,
provided that spatially resolved information on their [Fe/H]
abundances are available from IFU or long-slit spectroscopy. In the
context of the tidal stripping scenario, an important aspect in that
context will be to compare the CO index in nuclear regions of dwarf
galaxies with those in UCDs.

\acknowledgements {We thank the referee Michael Drinkwater for a
  very constructive report that improved the paper.} We are grateful
for fruitful discussions with Leopoldo Infante and J\"{o}rg Dabringhausen.
We thank Michael Hilker and Marina Rejkuba for
providing us with their data sets.



\begin{thebibliography}{}

\bibitem[2000]{Barmby00}Barmby, P., Huchra, J. P., Brodie, J. P. et al. 2000, AJ, 119, 727
\bibitem[1994]{Bassin94}Bassino, L. P., Muzzio, J. C., \& Rabolli, M. 1994, ApJ, 431, 634
\bibitem[2003]{Bekki03}Bekki, K., Couch, W.J., Drinkwater, M.J., Shioya, Y., 2003, MNRAS, 344, 399
\bibitem[2007]{Brodie07}Brodie, J. P., Cenarro, A. J., Beasley, M., Strader, J., \& Cardiel, N. 2007, NOAO Proposal ID \#2007A-0203
\bibitem[2003]{Bruzua03}Bruzual, G., \& Charlot, S. 2003, MNRAS, 344, 1000
\bibitem[2008]{Dabrin08}Dabringhausen, J., Hilker, M., \& Kroupa, P. 2008, submitted to MNRAS
\bibitem[2000]{Drinkw00}Drinkwater M.J., Jones J.B., Gregg M.D., Phillipps S.,
 2000, PASA 17, 227
\bibitem[2003]{Drinkw03}Drinkwater, M.J., Gregg, M.D., Hilker, M. {\it et~al.}, 2003, Nature, 423, 519
\bibitem[2004]{Drinkw04}Drinkwater M.J., Gregg M.D., \& Couch W.J. {\it et~al.} 2004, PASA, 21, 375
\bibitem[2007]{Evstig07}Evstigneeva, E. A., Gregg, M. D., Drinkwater, M. J., Hilker, M. AJ, 133, 1722
\bibitem[2002]{Fellha02}Fellhauer, M., \& Kroupa, P. 2002, MNRAS, 330, 642

\bibitem[2005]{Fellha05}Fellhauer, M., \& Kroupa, P. 2005, MNRAS, 359, 223
\bibitem[2006]{Fellha06}Fellhauer, M., \& Kroupa, P. 2006, MNRAS, 367, 1577
\bibitem[1978]{Frogel78}Frogel, J. A., Persson, S. E., Matthews, K., \& Aaronson, M. 1978, ApJ, 220, 75
\bibitem[2001]{Frogel01}Frogel, J. A., Stephens, A., Ramirez, S., \& DePoy, D. L. 2001, AJ, 122, 1896
\bibitem[2007]{Gilmor07}Gilmore, G., Wilkinson, M., I.; Wyse, R. F. G., Kleyna, J. T., Koch, A., Evans, N. W., Grebel, E. K. 2007, ApJ, 663, 948
\bibitem[1997]{Goldad97}Goldader, J. D., Joseph, R. D., Doyon, R., \& Sanders, D. B. 1997, ApJS, 108, 449
\bibitem[2005]{Hasega05}Ha\c{s}egan, M., Jord\'an, A., C\^{o}t\'{e}, P. {\it et~al.} (VCS team) 2005,  ApJ, 627, 203 
\bibitem[1999]{Hilker99}Hilker, M., Infante, L., Vieira, G., Kissler-Patig, M.,
   \& Richtler, T. 1999, A\&AS, 134, 75
\bibitem[2007]{Hilker07}Hilker, M., Baumgardt, H., Infante, L., Drinkwater, M., Evstigneeva, E., \& Gregg, M. 2007, A\&A, 463, 119
\bibitem[2000]{Ivanov00}Ivanov, V. D., Rieke, G. H., Groppi, C. E., Alonso-Herrero, A., Rieke, M. J., \& Engelbracht, C. W. 2000, ApJ, 545, 190

\bibitem[2006]{Jimene06}Jimenez, R. \& Haiman, Z. 2006, Nature, 440, 501
\bibitem[1994]{Kroupa94}Kroupa, P., \& Gilmore, G. F. 1994, MNRAS, 269, 655
\bibitem[2001]{Kroupa01}Kroupa, P., 2001, MNRAS, 322, 231
\bibitem[2003]{Kroupa03}Kroupa, P,, \& Bouvier, J. 2003, MNRAS, 346, 369
\bibitem[1992]{Lancon92}Lancon, A., \& Rocca-Volmerange, B. 1992, A\&AS, 96, 593
\bibitem[2005]{Marast05}Maraston, C. 2005, MNRAS, 362, 799
\bibitem[2005]{McLaug05}McLaughlin, D. E., van der Marel, R. P. 2005, ApJS, 161, 304
\bibitem[2007]{Metz07}Metz, M., \& Kroupa, P. 2007, MNRAS, 376, 387
\bibitem[2004]{Mieske04}Mieske, S., Hilker, M., Infante, L. 2004, A\&A, 418, 445
\bibitem[2006]{Mieske06}Mieske, S., Hilker, M., Infante, L., \& Jord\'an, A. 2006, AJ, 131, 2442
\bibitem[2007]{Mieske07}Mieske, S., Hilker, M., Jord\'{a}n, A., Infante, L., \& Kissler-Patig M. 2007, A\&A, 472, 111
\bibitem[2001]{Philli01}Phillipps S., Drinkwater M.J., Gregg M.D., Jones
  J.B., 2001, ApJ 560, 201
\bibitem[2007]{Rejkub07}Rejkuba, M., Dubath, P., Minniti, D., \& Meylan, G. 2007, A\&A, 469, 147
\bibitem[2007]{Tornat07}Tornatore, L., Ferrara, A., \& Schneider, R. 2007, MNRAS in press, arXiv:0707.1433
\bibitem[2007]{Walker07}Walker, M. et al. 2007, ApJ, 667L, 53
\bibitem[1994]{Worthe94}Worthey, G. 1994, ApJS, 95, 107
\bibitem[1988]{Zinnec88}Zinnecker, H., Keable, C. J., Dunlop, J. S., Cannon, R. D., \& Griffiths, W. K., 1988, Proceedings of the 126th Symposium of the International Astronomical Union,  Edited by Jonathan E. Grindlay and A. G. Davis Philip. Kluwer Academic Publishers, Dordrecht, 1988., p.603
\end{thebibliography}
\end{document}